\documentclass[prb,graphicx,preprint,showpacs,superscriptaddress]{revtex4-1}
\usepackage{amsmath,amssymb,amsthm,color}
\usepackage{graphicx}
\usepackage[english]{babel}
\usepackage{bm}

\begin{document}


\title{Chiral symmetry classes and Dirac nodal lines in three-dimensional layered systems} 



\author{Ching-Hong Ho}
\affiliation{Center for General Education, Tainan University of Technology, 710 Tainan, Taiwan}

\author{Cheng-Peng Chang}
\email[]{t00252@mail.tut.edu.tw}
\affiliation{Center for General Education, Tainan University of Technology, 710 Tainan, Taiwan}

\author{Ming-Fa Lin}
\affiliation{Department of Physics, National Cheng Kung University, 701 Tainan, Taiwan}


\date{\today}

\begin{abstract}
We study the existence and stability of Dirac nodal lines in three-dimensional layered systems, whose layers individually have Dirac nodal points protected by chiral (sublattice) symmetry. The model system we consider is the rhombohedral stack of graphene layers with each layer subjected to a uniform external potential that respects either AIII or BDI classes. From the Hamiltonians in either classes, a pair of nontrivial spiraling Dirac nodal lines can be derived. The results are reasonable in accord to the topological classification of gapless phases for codimension $2$. The nodal lines approach each other as the magnitude of the potential increases, revealing their annihilation due to the fact that regarding the full system their topological invariants are cancelled out.
\end{abstract}


\maketitle 

\par\noindent
Gapless phases of free fermions in condensed matters, including topological semimetals and nodal superconductors, have attracted much interest in recent years. A broad variety of relevant materials have been discovered or constructed today. Among others, a three-dimensional ($3$D) layered system can possibly have Dirac nodal lines (DNLs) in bulk momentum space if those layers individually have Dirac points (DPs) when decoupled. Hence, such DNLs are considered as continuous sets of DPs collected from the $2$D layers that are subjected to interlayer coupling. In the context of topological semimetals, for example, a dimensional crossover would occur, with a pair of spiraling DNLs emerging when graphene layers are stacked and coupled to $3$D rhombohedral graphite, where spinless Dirac fermions are hosted.\cite{heikkila11,ho16} Another example could be found in nodal superconductors with Majorana fermions.\cite{schnyder15}

A nontrivial Dirac node is robust against disorders or perturbations under the protection of a nonzero topological invariant.\cite{volovik07} The topological invariant is principally formulated according to the codimension of the node and the symmetry class to which its Hamiltonian belongs.\cite{zhao13} A nodal line in $3$D has co-dimension $p=3-1=2$ as a DP in $2$D. In connection to that for topological insulators/superconductors,\cite{schnyder08} a ten-fold classification of gapless phases has been established.\cite{zhao13,chiu14} As well, the three non-spatial symmetries, i.e., time-reversal symmetry (TRS), particle-hole symmetry (PHS) and chiral (sublattice) symmetry (SLS), interplay with the codimension. Both TRS and PHS are antiunitary while SLS is unitary as the combination of the former two. Of the ten classes, there is a single class  in which SLS is present even though TRS and PHS are absent. Although the presence of spatial symmetries might enrich the features of the matters,\cite{chiu14,koshino14} the ten-fold classification is complete in view of that unitary spatial symmetries are easy to break as referring, for example, to strained graphene.\cite{goerbig08,pereira09}

The origin of Dirac nodes depends on the system. In contrast to the PHS protected nodes in nodal superconductors,\cite{schnyder15} the DPs in $2$D graphene layers is protected by SLS.\cite{hatsugai11} An issue arising here is if one can learn the existence of DNLs from the decoupled $2$D layer by just considering the symmetry classes the $2$D Hamiltonians belong to. In other words, when the decoupled layer belonging to a specific SLS class has DPs, can one always obtains DNLs protected by the same class by stacking and coupling such layers? To clarify the above issue, we take a simplest model with external potential that respects different SLS classes.

Our model is based on the $3$D lattice shown in Fig.s 1(a) and (b), which is the rhombohedral stack (along the $z$ direction) of honeycomb-lattice layers [in the $(x,y)$ plane]. The bulk lattice consists of two sublattice as well as the $2$D honeycomb lattice. In fact, it can be taken as a $3$D extension of the Su-Schrieffer-Heeger (SSH) model,\cite{su79} which describes Dirac fermions about tremendous coupled chains of atoms.\cite{guinea06} Such systems would render non-trivial topology.\cite{xiao11,heikkila11} A realization of this model is given by rhombohedral graphite with graphene layers stacked in ABC configuration. The bulk Hamiltonian is written as
\begin{equation}
\mathcal{H}_{0}=-t\sum_{l}\sum_{\langle i,j\rangle}[a_{l,i}^{\dag}b_{l,j}+\mathrm{h.c.}]%
+t^{\prime}\sum_{l}\sum_{i}[b_{l+1,i}^{\dag}a_{l,i}+\mathrm{h.c.}],
\end{equation}
where $l$ labels the layers and $\langle i,j\rangle$ denotes nearest-neighbor sites on each layer, and the operators $a_{l,i}^{\dag}$ ($b_{l,i}^{\dag}$) create fermions in the sublattice $A$ ($B$) at $i$ site on $l$ layer. Owing to the absence of inter-sublattice coupling, there exists the chiral symmetry operator $\mathcal{S}=\sum_{l,\langle i,j\rangle}s_{l,\langle i,j\rangle}^{\dag}\sigma_{z}s_{l,\langle i,j\rangle}$, with $s_{l,\langle i,j\rangle}^{\dag}=(a_{l,i}^{\dag},b_{l,j}^{\dag})$, for $\mathcal{S}\mathcal{H}_{0}\mathcal{S}^{-1}=-\mathcal{H}_{0}$. The Fourier transformation by translation symmetry reads $a_{l,i}=N^{-\frac{1}{2}}\sum_{\mathbf{k}}e^{-i\mathbf{k}\cdot\mathbf{r}_{i,l}}a_{\mathbf{k}}$, and similar for $b_{\mathbf{k}}$, so that the Bloch Hamiltonian is given by
\begin{equation}
H(\mathbf{k})=d_{1}(\mathbf{k})\sigma_{1}+d_{2}(\mathbf{k})\sigma_{2},
\end{equation}
where the Pauli matrices $(\sigma_{1}, \sigma_{2}, \sigma_{3})$ act in the space of pseudospin $(a_\mathbf{k},b_\mathbf{k})$, and $d_{1}(\mathbf{k})=-t\sum_{m}^{3}\cos{(\mathbf{k}_{||}\cdot\mathbf{\delta}_{m})}%
+t^{\prime}\cos{(k_{z}c)}$ and $d_{2}(\mathbf{k})=t\sum_{m}^{3}\sin{(\mathbf{k}_{||}\cdot\mathbf{\delta}_{m})}%
-t^{\prime}\sin{(k_{z}c)}$ are obtained with $\mathbf{\delta}_{m}$ being the three vectors connecting nearest-neighbor sites and $c$ is the layer distance [Fig. 1(b)]. Thus, the chiral operator is given by $\mathcal{S}=\sigma_{z}$ and SLS is written as $\mathcal{S}^{-1}H(\mathbf{k})\mathcal{S}=-H(\mathbf{k})$ in momentum representation. It is easy to verify TRS by the operator $\mathcal{T}=\sigma_{0}\mathcal{K}$: $\mathcal{T}^{-1}H(\mathbf{-k})\mathcal{T}=H(\mathbf{k})$, and PHS by the operator $\mathcal{C}=\sigma_{z}\mathcal{K}$: $\mathcal{C}^{-1}H(\mathbf{-k})\mathcal{C}=-H(\mathbf{k})$, where $\sigma_{0}$ is the $2\times2$ identity matrix and $\mathcal{K}$ is the complex conjugation operator. Therefore, the Hamiltonian in Eq. (2) belongs to the chiral orthogonal BDI class.

The Hamiltonian described in Eq.s (1) and (2) is free from disorders and perturbations. There are three degrees of freedom ($k_{x}$, $k_{y}$, $k_{z}$) and two equations ($d_{1}=d_{2}=0$) for finding the zeros, which determine the existence of nodes. In rhombohedral graphite, a pair of DNLs are realized as shown in Fig. 2.\cite{mcclure69,ho13} The two DNLs spiral in opposite directions across the BZ boundaries from $k_{z}=-\pi$ to $k_{z}=\pi$. Now we apply an SLS-preserving external potential to each layer of this system. The simplest choice sufficing our purpose is to give a uniform (of course, also static) potential $w$ without Fourier components. Such potential is obviously invariant under transformation in the $C_{6v}$ group.\cite{manes07} Two cases are considered as follows.

$w=w_{1}\in \mathbb{R}$ \rule[1mm]{0.5cm}{0.2mm} The Hamiltonian is expressed in terms of
\begin{eqnarray}
d_{1}(\mathbf{k})&=&-t\sum_{m}^{3}\cos{(\mathbf{k}_{||}\cdot\mathbf{\delta}_{m})}%
+t^{\prime}\cos{(k_{z}c)}+w_{1}, \nonumber\\
d_{2}(\mathbf{k})&=&t\sum_{m}^{3}\sin{(\mathbf{k}_{||}\cdot\mathbf{\delta}_{m})}%
-t^{\prime}\sin{(k_{z}c)}.
\end{eqnarray}
It is also easy to verify that in the presence of this external real potential, the symmetry class of the $3$D layered system is still BDI as well as the $2$D decoupled layers. The DPs in each decoupled clean graphene layer for $w_{1}=0$ are well known to locate at the high-symmetry points $K^{(\pm)}=[\pm4\pi/(3a),0]$ [referring to Fig. 2]. For $w_{1}\neq0$, the DPs of each decoupled layer move in the $k_{x}$ direction according to $2\cos{(k_{x}a/2)}=-(1+w_{1})$, while leaving $k_{y}=0$ invariantly. The solvability gives the allowed range of perturbation, $-3\leq w_{1}\leq0$, where the DPs are located at
\begin{equation}
\mathbf{k}_{DP}^{(\pm)}=[\pm\frac{2}{a}\arccos{(\frac{-1-w_{1}}{2})},0].
\end{equation}
The two DPs meet and annihilate each other at $w_{1}=-3$. In order to find the zeros of Eq. (3) for DNLs, we transform $\mathbf{k}_{||}$ as $\mathbf{k}_{||}=\mathbf{k}_{DP}^{(\pm)}+\mathbf{k}$ and linearize $d_{1}$ and $d_{2}$ in Eq. (3) just for convenience. With polar coordinates $\mathbf{k}=(k,\phi)$, the two DNLs parameterized by $k_{z}$ are obtained as
\begin{eqnarray}
k_{DL}&=&\frac{t^{\prime}}{\hbar v}, \nonumber\\
\phi_{DL}&=&\mp(k_{z}c-\frac{\pi}{2})+\frac{\pi}{6},
\end{eqnarray}
where $v=\sqrt{3}at/(4\hbar)$ is the Fermi velocity. Equation (5) is in the same form as those not perturbed by $w_{1}$,\cite{ho13} except for the spiraling axes now being shifted to the $(\mathbf{k}_{DP}^{(\pm)},k_{z})$ lines.

We then derive the energy dispersion in terms of the polar
coordinates $(q, \theta)$ measured from the DPs for fixed $k{z}$. The coordinate
transformation runs as
$q^{2}=k^{2}+k_{D}^{2}-2k_{D}k\cos{(\phi-\phi_{D})}$ and
$\tan{\theta}=(k\sin{\phi}-k_{D}\sin{\phi_{D}})%
(k\cos{\phi}-k_{D}\cos{\phi_{D}})^{-1}$. Obviously,
$\theta\rightarrow -\theta$ as $\phi\rightarrow -\phi$ when changing
between the two valleys. The Dirac cone is thus derived
\begin{align}
\varepsilon(q, \theta)=\pm v\hbar q,
\end{align}
being independent of $k_{z}$.

$w=iw_{2}\in \mathbb{C}$, $w_{2}\in \mathbb{R}$, \rule[1mm]{0.5cm}{0.2mm} The imaginary external potential $iw_{2}$ results in a Hamiltonian
\begin{eqnarray}
d_{1}(\mathbf{k})&=&-t_{0}\sum_{m}^{3}\cos{(\mathbf{k}_{||}\cdot\mathbf{\delta}_{m})}%
+t^{\prime}\cos{(k_{z}c)}, \nonumber\\
d_{2}(\mathbf{k})&=&t_{0}\sum_{m}^{3}\sin{(\mathbf{k}_{||}\cdot\mathbf{\delta}_{m})}%
-t^{\prime}\sin{(k_{z}c)}-w_{2}.
\end{eqnarray}
The presence of $w_{2}$ breaks both TRS and PHS. However, SLS still holds and the Hamiltonian in Eq. (7) belongs to the chiral unitary AIII class as well as the $2$D decoupled layers do. The DPs of each decoupled layer should be found numerically.

In summary, we have shown that in $3$D layered systems DNLs can be protected by the same SLS class as the DPs in those decoupled layer are. It should be remarked that the annihilation of the pair of DNLs cannot be fulfilled even though the two spiraling axes coincide. To show the annihilation, other types disorders or perturbations regarding fermion doubling might be included, which must involve the $z$ dimension.

\begin{acknowledgments}
This work was supported by the Ministry of Science and Technology of Taiwan,
under the Grant nos. MOST 105-2811-M-165-001 and MOST 105-2112-M-165-001-MY3.

\end{acknowledgments}

\newpage

\bigskip \vskip0.6 truecm\newpage
\centerline {\Large \textbf {Figure
Captions}}
\begin{itemize}

\item[FIG. 1.](Color online) (a) Rhombohedral lattice of $3$D stack of $2$D honeycomb lattices, where the rhombohedron (red) is the biparticle primitive unit cell. The present minimal model is described in terms of the intralayer hopping $t$ and interlayer hopping $t^{\prime}$. (b) Honeycomb lattice of each $2$D layer. The nearest-neighbor sites associated with hopping $t$ are connected by three vectors $\mathbf{\delta}_{m}$. (c) Schematic of the $3$D extension of SSH model constructed by a rhombohedral stack of $2$D honeycomb-lattice layers. One representative chain is shown by linked thick sticks where intralayer hopping $t$ (blue) and interlayer hopping $t^{\prime}$ (yellow) take place.

\item[FIG. 2.](Color online) (a) Projections (red circles) of the spiraling DNLs on the $2$D projected BZ (blue hexagon), where the portions of DNLs inside (solid) and outside (dotted) are shown. To sum up, there are two inequivalent DNLs in the $3$D rhombohedral BZ. The arrows indicate the spiraling senses in the increase of $k_{z}$. (b) $3$D rhombohedral BZ (red), with the unfilled dots on the high-symmetry points, in company with the $2$D projected BZ.
\end{itemize}


\begin{figure}[p]
\includegraphics[scale=0.8]{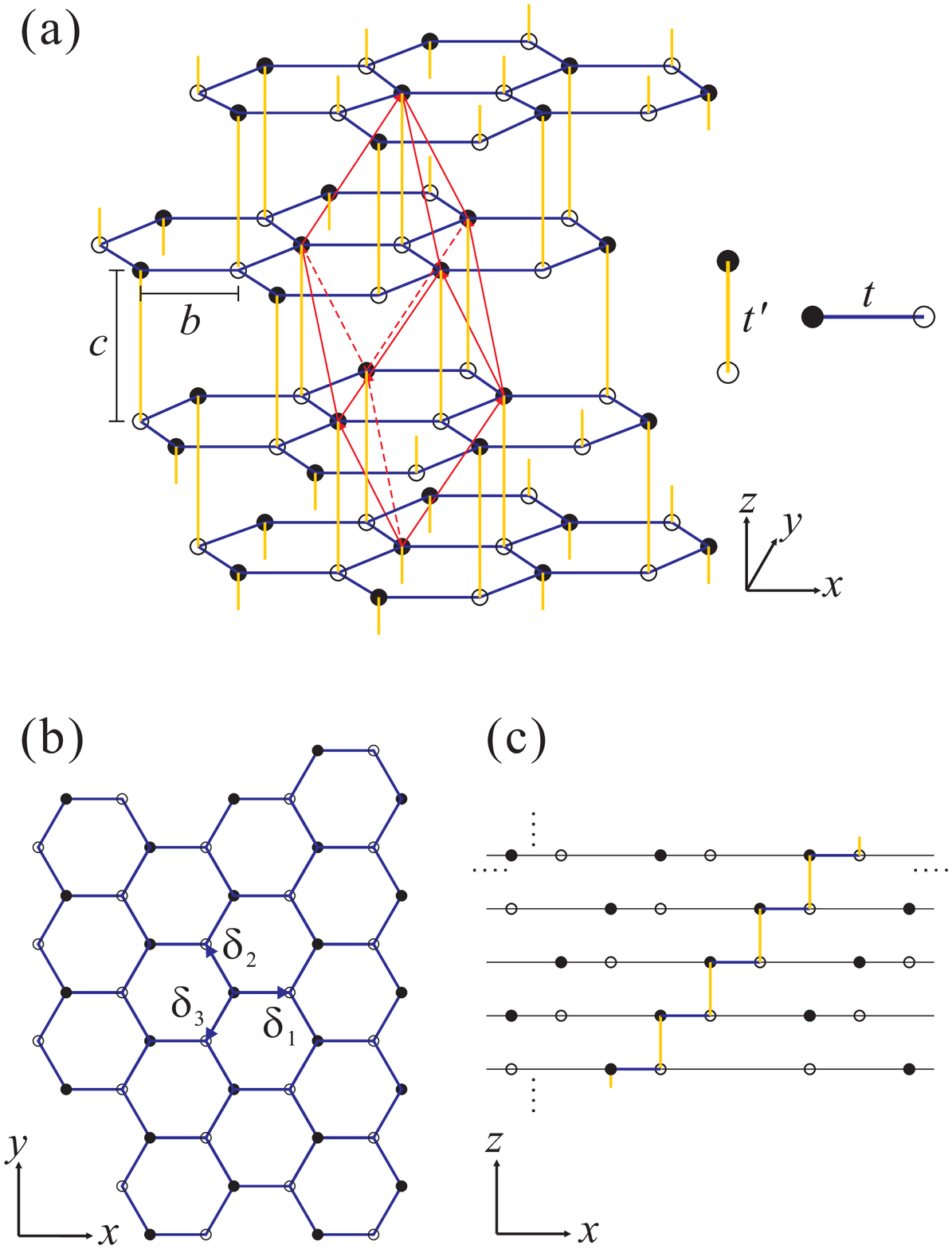}
\end{figure}

\begin{figure}[p]
\includegraphics[scale=0.8]{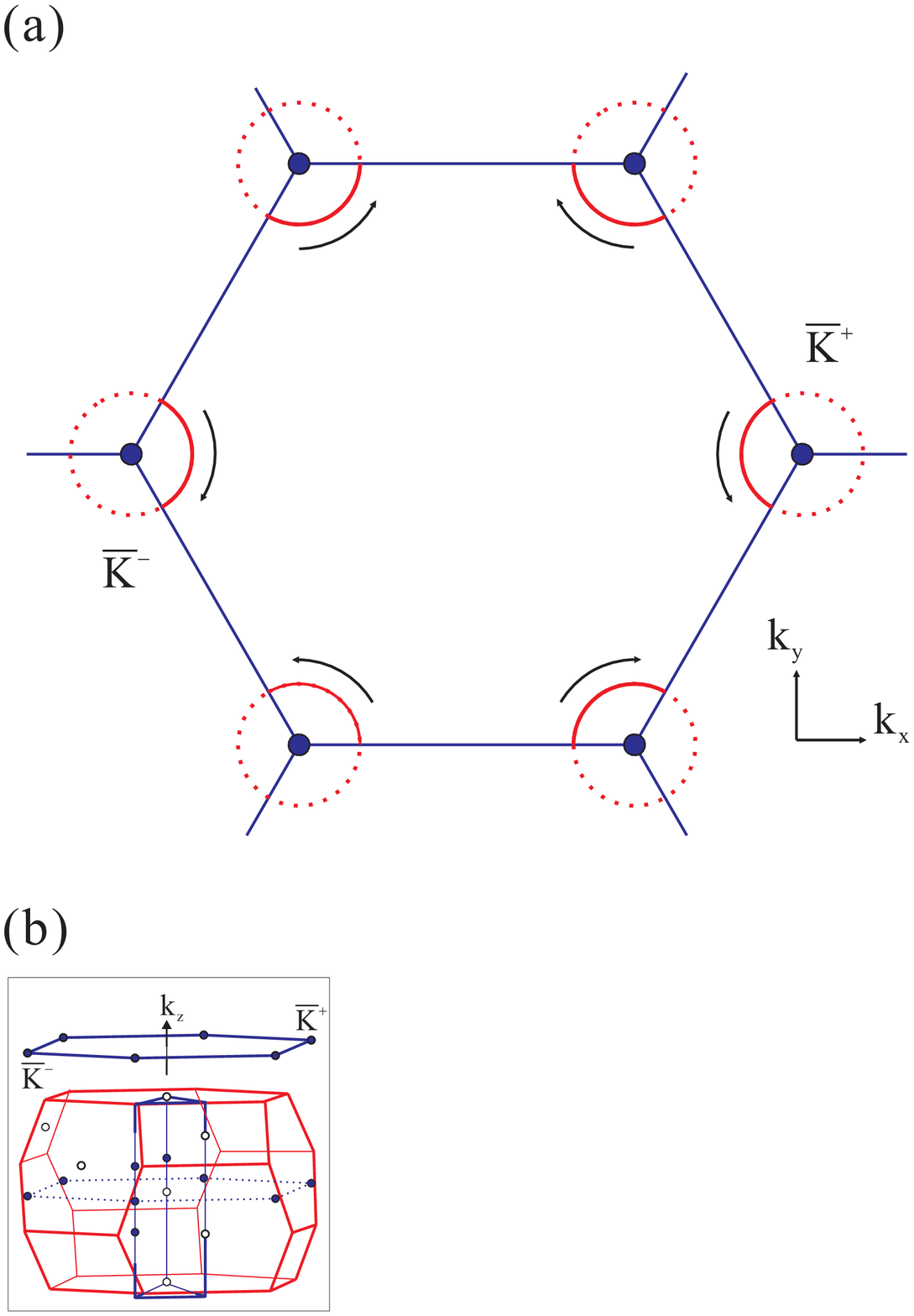}
\end{figure}

\end{document}